\documentclass[pra,aps,twocolumn,showpacs,amsmath,amssymb,floatfix]{revtex4}

\usepackage{graphicx}
\usepackage{dcolumn}
\usepackage{bm}
\usepackage{epsfig}
\usepackage{amsmath}

\newcommand{\be}{\begin{equation}}
\newcommand{\ee}{\end{equation}}
\newcommand{\beq}{\begin{eqnarray}}
\newcommand{\enq}{\end{eqnarray}}

\begin{document}

\title{Dipolar Drag in Bilayer Harmonically Trapped   Gases}

\author{N. Matveeva, A. Recati and S. Stringari}
\affiliation{Dipartimento di Fisica, Universit\`a di Trento and INO-CNR BEC Center, I-38123 Povo, Italy}
 
\begin{abstract}
We consider  two separated pancake-shaped trapped gases interacting with a dipolar (either magnetic or electric) force. We study how the center of mass motion propagates from one cloud to the other as a consequence of the long-range nature of the interaction. The corresponding dynamics is fixed by  the frequency difference between the 
in-phase and the out-of-phase center of mass modes of the two clouds, whose dependence on the dipolar interaction strength and the cloud separation is explicitly investigated. 
We discuss  Fermi gases in the degenerate as well as in the classical limit and comment on the case of Bose-Einsten condensed gases.
\end{abstract} 


\maketitle

\section{Introduction}

In recent years atomic and molecular dipolar gases have attracted a lot of interest since the long-range and the anisotropic nature of the interaction is expected to give rise to
new important features at both the microscopic and macroscopic level. These  include, among others,   novel effects in the mechanism of the expansion and of the collective modes  \cite{collective} and in the structure of the superfluid \cite{SFdipole} and normal \cite{LFtheory} phase, new exotic phases of crystalline nature (see e.g., the recent work \cite{exotic}) and new schemes for  quantum computation in the presence of optical lattices \cite{QC}. Some of these effects, in particular those concerning the expansion and the collective modes, have been already experimentally  observed in magnetic dipolar atomic gases\cite{Pfau}. The recent progress  in the realization of gases of electric polar molecules, where the effect of the dipolar force is particularly strong, is expected to open new challenging frontiers in this area of research (see \cite{RevNJP,dipmolSilke, dipmolDeMille} and references therein).

\begin{figure}[t]
\includegraphics[height=7cm]{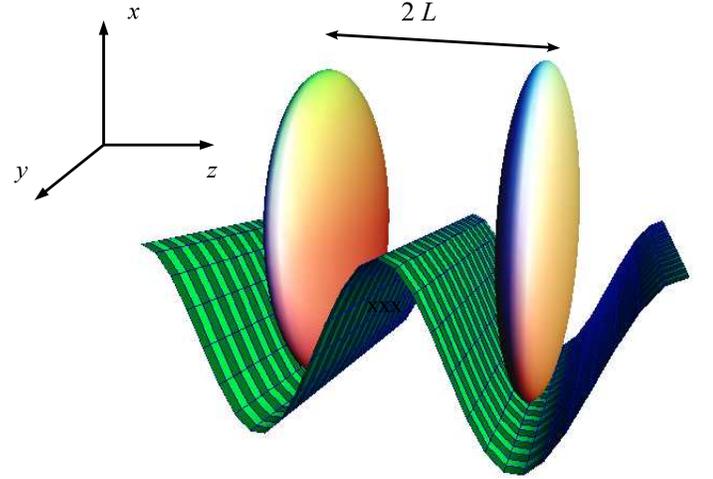}
\caption{Scheme of two not overlapping pancake shaped clouds of a dipolar  gas. The distance between the centers of mass of the clouds is $2 L$. The clouds are harmonically confined in the transverse directions $x$, $y$.}
\label{fig:sketch}
\end{figure}

The aim of the present work is to propose a drag experiment induced by the long range nature of the dipolar interaction. We consider an atomic or molecular  gas harmonically trapped in a double well configuration such that the  overlap  between the two clouds and the corresponding tunneling effect can be neglected   (see Fig. \ref{fig:sketch}). The only force acting between the two gases is of long range nature (here and in the following we assume that dipoles are oriented in the direction orthogonal to the discs, i.e along the $z$-th axis of Figure 1) and we study how the out-of-phase transverse dipole mode is affected by the long-range interaction. Displacing one of the two clouds out of its equilibrium position and releasing it, will excite both the in-phase (center of mass) and the out-of-phase dipole modes. On  a time scale fixed by the inverse of the frequency difference between the two modes, the center of mass motion of the first cloud will be transferred to the second one. We call this effect ``dipolar drag" in analogy to the well known  Coulomb drag (see e.g., \cite{Rojo}) exhibited by  electrons in uniform bilayer systems\cite{noteSarmaDrag}.

\section{Dipolar Drag of the center-of-mass motion}

\begin{figure}[t!]
\center{
\includegraphics[height=5.8cm]{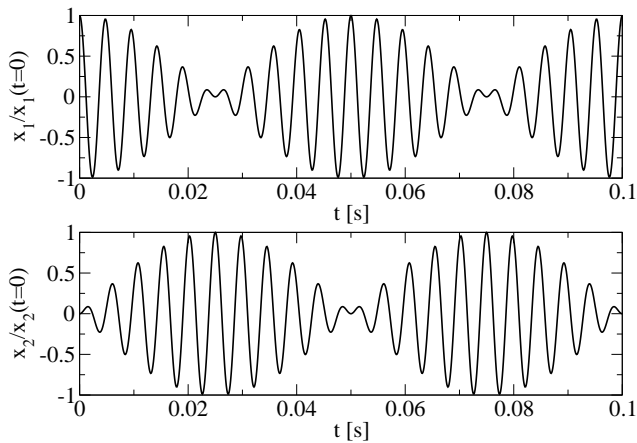}
}
\caption{The motion of the center of mass of the two clouds for $\omega_{out}=1.1\omega_\perp$, with $\omega_\perp/2\pi=200$ Hz which corresponds to a beating time ${\bar t}=\pi/(\omega_{out}-\omega_\perp)=0.025$s. Initially, at $t=0$, only the cloud $1$ is displaced from its central equilibrium position. }
\label{fig:osc1}
\end{figure}

We consider a gas confined by a cylindrically harmonic potential:  
\begin{equation}
V^{1,2}_{trap}(x, y, z)=\frac{1}{2}m\omega_\perp^2[x^2+y^2+\lambda^2 (z\pm z_0)^2].
\label{eq:Vext}
\end{equation}
where $2 z_0$ is the distance between the minima of the potential along $z$, $\lambda=\omega_z/\omega_\perp$ is the ratio between the transverse and longitudinal trapping frequencies and we consider 
pancake configurations, i.e., $\lambda\gg 1$. 
Let $x_i$ being the center of mass coordinate along $x$ of the $i$-th cloud. The equations of motion can be written as 
\begin{eqnarray}
\frac{d x_1}{dt}&=&\omega_\perp^2 x_1+\alpha (x_1-x_2),\\
\frac{d x_2}{dt}&=&\omega_\perp^2 x_2-\alpha (x_1-x_2),
\label{coupHO}
\end{eqnarray}
where $\alpha$ is the coupling between the two bare center-of-mass modes. The eigenfrequencies of the previous equation are simply $\omega_{in}=\omega_\perp$, for the in-phase sloshing mode and $\omega_{out}=\omega_\perp\sqrt{1+2\alpha/\omega_\perp^2}$ for the  out-of-phase sloshing mode. Thus in order to determine $\alpha$ we just need to determine the splitting $\omega_{out}-\omega_\perp$ for the dipolar coupled system.
Once the frequency $\omega_{out}$ is known, we can determine quantitatively the evolution of the system as described by Eq. (\ref{coupHO}).  In Fig. \ref{fig:osc1} the motion of the coupled clouds for a value of $\omega_{out}=1.1\omega_\perp$ (see Sec. \ref{sec:III} below) is shown. The beating of the motion is a direct measurement of the out-of-phase mode frequency, since the time at which the initially displaced cloud stops in the center is simply ${\bar t}=\pi/(\omega_{out}-\omega_\perp)$. 

In the following we calculate the frequency $\omega_{out}$ as a function of the dipolar interaction strength and the distance between the two clouds.
We will also discuss how the equation of state of the gas  affects such a frequency. The frequency $\omega_{out}$ was recently calculated by  Huang and Wu \cite{HuangWu} in the case of a magnetic dipolar Bose gas, using a technique very similar to the one employed in the present work. For this reason we mainly focus on the case of  Fermi gas. 
Moreover the Fermi statistics allows for an easier realization of cold gases of  hethero-nuclear molecules carrying an electric dipole moment (e.g., the recent experiment \cite{StereoYe}) so that the strength of the dipolar force can be much larger. 

\section{Two coupled Fermi gases with dipolar interaction}
\label{sec:III}

The system we want to study is made of two clouds, $i=1,2$, of dipolar Fermi gas in the normal phase confined by the 
potential Eq. (\ref{eq:Vext}).
Moreover we assume that the dipoles are oriented along  the $z$ axis by an electric field. 
The interaction potential between two dipoles ($\vec{d_1}=\vec{d_2}=\vec{d}$) has the standard form:
\begin{equation}
V_D(\vec{r_1},\vec{r_2},\theta)=\frac{d^2 (1-3 \cos^2 \theta )}{|\vec{r_1}-\vec{r_2}|^3}, \\
\label{eq:VD}
\end{equation}
where $\theta$ is the angle between $\vec{d}$, i.e., the $z$ direction, and $\vec{r_1}-\vec{r_2}$.

We study the system by means of the following energy functional of the cloud densities $n_i$, $i=1,2$
\begin{eqnarray}
E[n_1,n_2]&=&\sum_{i=1,2}(E^i_{kin}[n_i]+E^i_{trap}[n_i]+E^{ii}_{dd}[n_i])\nonumber\\
&+&E^{12}_{dd}[n_1,n_2], \label{eq:Etot}
\end{eqnarray}
where we introduce the kinetic energy
$E^i_{kin}=\hbar^2/(20 m\pi^2)\int{d\vec{r}[6\pi^2n_i(\vec{r})]^{5/3}}$ calculated within local density approximation, the potential energy $E^i_{trap}=\int{d\vec{r}[n_i(\vec{r})V^i_{trap}]}$ and the intra- and inter-cloud dipolar energies
\be
E^{ij}_{dd}=\frac{1}{2}\int{n_i(\vec{r})n_j(\vec{r'})V_D(\vec{r}-\vec{r'})d\vec{r}d\vec{r'}}.
\ee
In the energy functional Eq. (\ref{eq:Etot}) we have not included the intra-cloud exchange energy. We safely neglect it, since for the pancake-like configurations, which we are mainly interested in, the direct term is the dominant effect (see, e.g., \cite{Miyakawa}). 

In order to study the center-of-mass oscillations in the transverse direction (see Fig. \ref{fig:sketch}) we consider a scaling transformation for the densities of the type
$
n_i(x,y,z)\rightarrow n_i(x+\epsilon_i,y,z).
$
From the variation of the energy functional (\ref{eq:Etot}) we get as expected two modes. The in-phase mode is not affected by the dipolar interaction and has a frequency $\omega=\omega_\perp$ equal to the harmonic trapping one. Conversely the out-of-phase mode is affected by the dipole interaction and is characterized by the frequency
\be
\frac{\omega_{out}}{\omega_\perp}\!=\!\!\sqrt{1-\!\!\frac{2}{m\omega_\perp^2N}\!\!\int\!\!{d\vec{r}_1 d\vec{r}_2} V_D(\vec{r}_1-\vec{r}_2) \frac{\partial n_1(\vec{r}_1)}{\partial x} \frac{\partial n_2(\vec{r}_2)}{\partial x}},
\label{eq:omegaD}
\ee 
where $N$ is the total atom/molecule number.

Assuming the two clouds are identical we can get the densities of the clouds by means of an easy variational gaussian ansatz 
\be
n_{1,2}(x,y,z)=\frac{N \kappa}{W_\perp^3 \pi^{\frac{3}{2}}} \exp\left(-\frac{x^2+y^2+\kappa^2(z\mp L)^2}{W_\perp^2}\right), \label{eq:anzatz}
\end{equation}
where $W_\perp$ and $W_z=W_\perp/\kappa$ are the widths of single cloud and $2 L$ is the distance between the clouds, which, for our parameters, namely a strong confinement in the $z$ direction, is very close to the distance $2 z_0$.

Inserting the gaussian anzatz Eq. (\ref{eq:anzatz}) into the single cloud energy one obtains  
\begin{eqnarray}
&&E(W_\perp,\kappa)=\left(\frac{3}{5}\right)^{5/2}\frac{\hbar^2 N}{2m\pi}\left(\frac{6\pi^2 N\kappa}{W_\perp^3}\right)^{2/3}\nonumber\\
&&+\;mN\omega_\perp^2 W_\perp^2\left(1+\frac{\lambda^2}{2\kappa^2}\right)-\frac{d^2N^2 \kappa}{3\sqrt{2\pi}W_\perp^3}f(\kappa),\label{eq:Etot1}
\end{eqnarray}
where we introduce the standard notation (see e.g. \cite{chiara})
\begin{equation}
f(\kappa)=\frac{1+2\kappa^2}{1-2\kappa^2}-\frac{3\kappa^2\arctan{\sqrt{1-\kappa^2}}}{(1-\kappa^2)^{3/2}}.\nonumber
\end{equation}	

\begin{figure}[t]
\centering
\includegraphics[height=5.5cm]{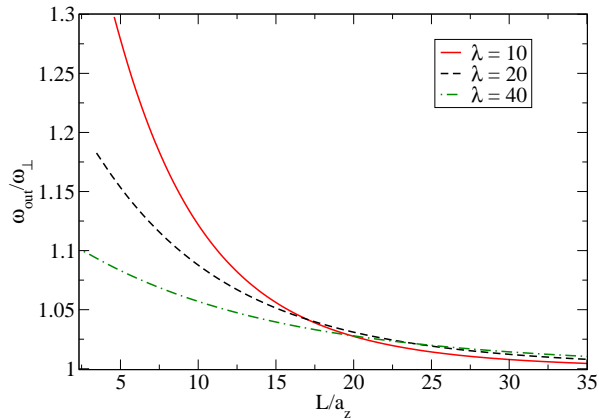}
\caption{Frequency shift of $\omega_{out}/\omega_\perp$ for the out-of-phase mode for a degenerate gas of $^{40}$K$^{87}$Rb, with parameters as given in Table \ref{table:nonlin}.}
\label{fig:wD}
\end{figure}

Once the densities are known we are in the position to calculate the frequency $\omega_{out}$.
The problem has many parameters and to be concrete we consider a gas of bi-atomic  molecules $^{40}$K$^{87}$Rb and reasonable experimental values for the number of molecules and for the trapping potentials (see Table \ref{table:nonlin}).
The result for the out-of-phase mode frequency as a function of the clouds' distance and for different value of $\lambda$ are reported in Fig. \ref{fig:wD}. 

\begin{table}[t]
\caption{ The cloud  size for $N=2200$ $^{40}$K$^{87}$Rb molecules with $\omega_z/2\pi=10$ kHz (corresponding to $a_z=8.89\times10^{-6} $cm) and dipole momentum  $d=0.56 $ D.} 
\vspace{0.1cm}
\centering 
\begin{tabular}{| c | c | c | c | c |} 
\hline 
$\lambda$ & $W_\perp$, $10^{-4}$cm & $W_z$, $10^{-5}$cm  \\ [0.5ex] 
\hline 
10 & 1.8  & 2.05  \\
20 & 2.9  & 1.54 \\
40 & 4.6  & 1.17 \\[1ex] 
\hline 
\end{tabular}
\label{table:nonlin} 
\end{table}

We see that for small enough distance the larger the cloud (larger $\lambda$ for fixed $N$) the smaller the effect.
This can be easily understood in terms of the potential of a single disk of radius $W_\perp$ on a probe dipole. Indeed at a distance $z\ll W_\perp$ the potential decays with $W_\perp$, which
is a general result independent of statistics.
On the other hand we have that the asymptotic behavior of the frequency shift at large distance is $\propto \sqrt{1+C/L^5}$ with $C$ a constant, which 
is the result one immediately obtaines by considering just two trapped dipoles. It can also be easily shown that considering spherical clouds with $W_\perp=W_z=W$ in Eq. (\ref{eq:omegaD}) the frequency of out-of-phase dipole mode reads
\begin{equation}
\omega_{out}=\omega_\perp\left(1-\frac{\sqrt{2} Nd^2h(L/W)}{3\sqrt{\pi} m\omega_\perp^2L^5}\right)^{1/2},\label{eq:shift_sp}
\end{equation}
where  $h(y)=e^{-2y^2}(4y^5+6y^3+9/2y)-9/2\sqrt{\frac{\pi}{2}}{\rm Erf}(\sqrt{2}y)$, which 
approaches a constant for large values of $y$. 

The results reported in Fig. \ref{fig:wD} show how the frequency shift can be large enough, for the chosen parameters, to be experimentally measurable. 
Moreover we checked that the results are the essentially the same using Thomas-Fermi profiles, instead of the gaussian ansatz Eq. (\ref{eq:anzatz}).

It is useful to compare the above predictions for the frequency shifts
calculated for a zero temperature Fermi gas with the ones holding for  a Bose-Einstein
condensed gas or for a classical thermal configuration. To this purpose
one can still use Eq. (\ref{eq:omegaD}) with the proper density profiles (an inverted
parabola for  a BEC gas and a Boltzmann distribution for a classical gas).
As emerges from Eq. (\ref{eq:omegaD}) the effect is amplified for smaller radial sizes 
where the gradient of the density is larger. One then understands that a
Bose gas interacting with a moderate value of the scattering length will
provide larger shifts with respect  to both a Fermi gas and a thermal
configuration.  The shifts for a Bose gas were investigated in \cite{HuangWu} where,
however, magnetic dipolar atomic gases were considered, which are
characterized by a significantly small value of the dipolar coupling constant $d^2$ in Eq. (\ref{eq:VD}).  
At present the more promising perspectives for
realizing electric dipolar molecules, where the effect of the dipolar
interaction is particularly strong, concern the fermionic  species which
ensure better  stability conditions and which are already available in the
thermal regime.
\begin{table}[t]
\caption{The size of cloud for classical gas of $^{40}$K$^{87}$Rb at $T= T_F$ with dipole momentum  $d=0.56 $D and trapping frequency 
$\omega_z=10$ kHz ($a_z=8.89\times10^{-6} $cm)).} 
\vspace{0.1cm}
\centering 
\begin{tabular}{|c| c| c| c| c| } 
\hline 
$\lambda$ & $W_\perp$, $10^{-4}$cm & $W_z$, $10^{-5}$cm  \\ [0.5ex] 
\hline 
10  & 2.8  & 2.8  \\
20  & 4.5  & 2.25 \\
40  & 7.15 & 1.79 \\ [1ex] 
\hline 
\end{tabular}
\label{table:T} 
\end{table} 
\begin{figure}[]
\centering
\includegraphics[height=5.5cm]{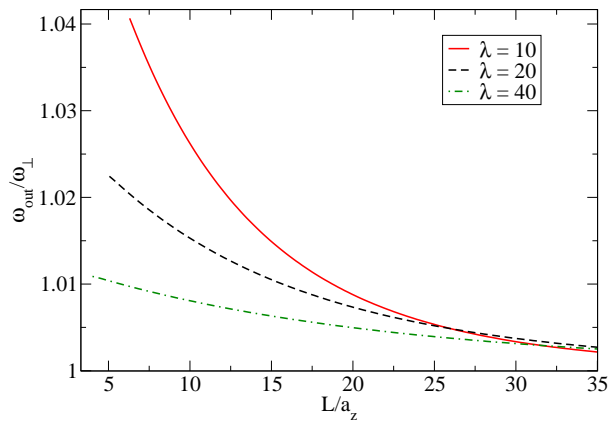}
\caption{Frequency shift of $\omega_{out}/\omega_\perp$ for the out-of-phase mode for a classical gas of $^{40}$K$^{87}$Rb at temperature $T=T_F$, where $T_F$ is the Fermi temperature of the gas. Parameters are given in Table \ref{table:T} .}
\label{fig:boltz}
\end{figure}

In Fig. \ref{fig:boltz} we report the predictions for the frequency
shifts exhibited by  a thermal configuration calculated at the temperature $T=T_F$, where $T_F\equiv \hbar{\omega_\perp}(6N\lambda)^{1/3}/k_B$ is the Fermi energy and $k_B$ the Boltzmann's constant. We used simply the gaussian density profiles of Eq. (\ref{eq:anzatz}), but with 
the radii given by the Boltzmann expression $W_\perp^2=2k_B T/(m\omega_\perp^2)$ and $\kappa=\lambda$.
The corresponding parameters
are given in Table \ref{table:T}. Comparison with the predictions reported in Fig. \ref{fig:wD}  shows that
the effect, for the same trapping conditions and number of particles, is
indeed smaller than for  a degenerate Fermi gas, since the thermal radii are larger and the densities smaller  
than the ones of the degenerate configuration. 

\section{Conclusions}
We have proposed a drag experiment between two non-overlapping atomic/molecular clouds (see Fig. \ref{fig:sketch} and Fig. \ref{fig:osc1}) to test the long-range nature of the dipolar potential. The method is 
independent of quantum statistics and holds for both degenerate and thermal gases. This effect corresponds to the trapped version of the famous Coulomb drag exhibited by electrons in uniform bilayer systems. The realization of such a drag experiment would provide a direct and easy signature of the long-range nature of the dipole interaction.

\section*{Acknowledgement}
Useful discussions with M. Abad, M. Klawunn, S. Ospelkaus and L. P. Pitaevskii are ackowledged.
This work has been supported by ERC through the QGBE grant.

\end{document}